\documentclass[a4paper,11pt]{article}
\usepackage{jheppub}
\usepackage{lineno}
\usepackage{graphicx,amsmath,hyperref,amsfonts}
\usepackage{pgfplots}
\usepackage{orcidlink}
\pgfplotsset{compat=1.18}
\usepackage{tikz}
\renewcommand{\d}{{\rm d}}
\newcommand{\nn}{\nonumber}

\title{\boldmath Holographic free energy and integrated correlators of ${\cal N}=1^*$ theory}

\author{Nakwoo Kim\,\orcidlink{0000-0002-4182-8459}}
\author{and Hoseob Shin\,\orcidlink{0009-0005-5662-2399}}
\affiliation{Department of Physics and Research Institute of Basic Science, Kyung Hee University, \\  26, Kyungheedae-ro, Dongdaemun-gu,
Seoul 02447, Republic of Korea}

\emailAdd{nkim@khu.ac.kr}
\emailAdd{hoseob0610@khu.ac.kr}

\abstract{We consider the 2nd integrated correlators of ${\cal N}=4$, $D=4$ super Yang-Mills theory, especially those which can be associated with ${\cal N}=1^*$ mass-deformed theories. We provide an analytic derivation of the integrals at supergravity tree level, which has not been available so far. Our result agrees with the previous results from the study of BPS solutions in supergravity and the holographic evaluation of the sphere partition function.}

\arxivnumber{2411.08615}

\begin{document}
\maketitle
\flushbottom

\section{Introduction}
\label{sec:intro}
A lot of effort has been devoted to thorough understanding of ${\cal N}=4$, $D=4$ super Yang-Mills (SYM) theory, especially as the prototype example of AdS/CFT correspondence \cite{Maldacena:1997re}. Recently we have witnessed considerable progress made on the topic of integrated correlators \cite{Chester:2020dja,Chester:2020vyz}. Here we are interested in the 4-point correlators of 1/2-BPS operators. They in general have non-trivial coordinate dependence in terms of cross ratios, but we are specifically interested in the quantity of 4-point correlators, integrated with an appropriate measure which is consistent with the superconformal invariance. The resulting quantities can be interpreted as certain 4th derivatives of the free energy $\log Z(\tau,\bar \tau, m)$ for generally mass-deformed SYM with mass $m$ and complexified coupling constant $\tau$. Using the common terminology, the 1st integrated correlators are $\partial_\tau \partial_{\bar \tau} \partial^2_m\log Z(\tau,\bar \tau, m)|_{m=0}$. By the 2nd integrated correlators we mean $\partial^4_m \log Z|_{m=0}$. A large portion of the literature so far has considered the so-called ${\cal N}=2^*$ theory, where the ${\cal N}=4$ SYM matter content is split into a ${\cal N}=2$
vector multiplet and a ${\cal N}=2$ hypermultiplet, and we assign a non-zero mass to the hypermultiplet to break ${\cal N}=4$ supersymmetry to ${\cal N}=2$. The advantage of this particular deformation is that we can make use of the powerful machinery of supersymmetric localization \cite{Pestun:2007rz}, which enables us to reduce the path integral of certain physical observables to a matrix model. Integrated correlators have been analyzed using string/graviton scattering amplitudes \cite{Drummond:2020uni,Bissi:2020woe,Abl:2020dbx}, using bootstrap \cite{Bissi:2020jve,Alday:2021vfb}, and using integrability \cite{Cavaglia:2022yvv,Cavaglia:2023mmu}. 

Let us give an outline and a brief summary of the progress made on this topic so far. Modular invariance of integrated correlators and its lattice-sum representation involving Eisenstein series has been 
conjectured in \cite{Dorigoni:2021bvj,Dorigoni:2021guq}. In a very interesting work, \cite{Collier:2022emf} exploited S-duality of ${\cal N}=4$ SYM to determine the gauge coupling dependence of certain CFT observables including the integrated four-point functions at both
perturbative and non-perturbative level. The standard perturbative computations of integrated 4-point correlators have been performed and the results are summarized in terms of elementary Feynman graphs in \cite{Wen:2022oky}. In \cite{Hatsuda:2022enx}, the authors gave evidence on the interpretation of exact formula for integrated correlators, given in \cite{Dorigoni:2022zcr}, as sum over $(p,q)$-strings and the non-perturbative corrections as D3-brane instantons. Integrated correlators have been also studied using bootstrap method in \cite{Caron-Huot:2022sdy}. The modular invariance of the 2nd integrated correlators and its strong coupling regime was studied in \cite{Paul:2022piq}. A lattice sum representation of the 1st integrated correlator and in particular its modular invariance has been elucidated in \cite{Dorigoni:2022cua}. A localization computation of integrated 4-point correlation functions in the planar limit and perturbative expansion of 't Hooft coupling constant has been performed in \cite{Fiol:2023cml}. Perturbative computations of integrated 4-point correlators, which are not accessible by supersymmetric localization, have been reported in \cite{Brown:2023zbr} in the planar limit and to all orders in 't Hooft coupling.  
Using localization, \cite{Alday:2023pet} developed a method to compute the second integrated correlator in large-$N$ expansion up to order $1/N^3$ and proposed 
certain properties of integrated correlators which are valid to all orders in 
$1/N$. 
A recent reference \cite{Pini:2024uia} studies the second integrated correlator for ${\cal N}=2$ theory obtained via orbifolding of ${\cal N}=4$ SYM. They use localization formulae and also consider perturbative computations. More recent works on (integrated) 4-point correlators can be found in e.g. \cite{Zhang:2024ypu,Billo:2024ftq,Loebbert:2024fsj,Turton:2024afd,Aprile:2024lwy,Dorigoni:2025beu,Chester:2025kvw,DeLillo:2025stg,Brown:2025huy,Brown:2026dhy,Aprile:2026urq,Chester:2026uro}, in particular including integrated correlators with line defects \cite{DeLillo:2025hal,DeLillo:2025eqg,DeLillo:2026gyy} and in massive ${\cal N}=2$ theories \cite{Damia:2025eed}. See also \cite{Bobev:2025gzu} for holographic correlators of the ${\cal N}=1$ Leigh-Strassler SCFT, \cite{BenettiGenolini:2025icr} for equivariant localization of $D=5$ gauged supergravity, and \cite{Karydas:2025tfs,Derkachov:2025xmt} for recent developments on conformal ladder integrals.

The goal of this work is twofold. First, among the various integrated correlation function calculations in the seminal paper \cite{Chester:2020dja}, there is a particular two-dimensional integration which is supposed to give the 4th mass derivative of the ${\cal N}=2^*$ free energy at tree level supergravity. To be consistent with the localization computation in large $N$ limit and also with supergravity computation \cite{Bobev:2013cja}, this integral (Eq.\eqref{ic2star} in this article) must give $3-6\zeta(3)$ but it was only numerically checked in \cite{Chester:2020dja}. We provide an analytic calculation here. Second, there exists a holographic model which is dual to ${\cal N}=1^*$ theory \cite{Bobev:2016nua}, and although there is no hint from supersymmetric localization, the 4th mass derivative of the holographic free energy has been analytically given in \cite{Kim:2019rwd}. One naturally expects for these ${\cal N}=1^*$ data to be reproducible as a certain sum of integrated 4-point correlation functions between scalar and spinor mass terms, just like the ${\cal N}=2^*$ case. We show it is indeed possible, building on the ingredients of 4-point functions given as derivatives of the scalar ladder diagram \cite{Usyukina:1993ch}.
The plan of this work is as follows. In Sec.\ref{sec:2} we review the results of holographic computation based on the fully-backreacted $D=5$ supergravity solutions. Sec.\ref{sec:3} is the main part and we provide an analytic method to compute the integrals relevant to the 2nd integrated correlators. We conclude with discussions in Sec.\ref{sec:4}. A Mathematica notebook reproducing the results of Sec.~\ref{sec:3} is provided as an ancillary file with the arXiv version of this article.
\section[Holographic free energy of mass-deformed SYM]{Holographic $S^4$ free energy of ${\cal N}=1^*$ 
and ${\cal N}=2^*$ theories}
\label{sec:2}
The subsector of Wick-rotated ${\cal N}=8$ gauged supergravity in $D=5$, relevant to ${\cal N}=1^*$ mass-deformation of ${\cal N}=4$ super Yang-Mills theory on (unit-radius) $S^4$ has been presented along with the associated BPS equations and numerical solutions in \cite{Bobev:2016nua}. The model has, in addition to the Einstein-Hilbert term, 10 real scalar fields whose interactions are described in terms of superpotential and K\"ahler potential. The Lagrangian is 
\begin{align}
{\cal L}&=-\frac{1}{4}R + 
3 \frac{(\partial{\eta_1})^2}{\eta_1^2}
+\frac{(\partial{\eta_2})^2}{\eta_2^2} + \frac{1}{2}{\cal K}_{a \bar{b}} \partial_{\mu} z^a \partial^{\mu} \bar{z}^{\bar{b}} -{\cal P} ,
\label{sugra5}
\end{align}
The scalar potential ${\cal P}$ is given using a superpotential ${\cal W}$,
\begin{align}
{\cal P}& = \frac{1}{8} e^{\cal K} \left(\frac{\eta_1^2}{6} \partial_{\eta_1}{\cal W} \partial_{\eta_1}{\cal \widetilde{W}}+\frac{\eta_2^2}{2} \partial_{\eta_2}{\cal W}\partial_{\eta_2}{\cal \widetilde{W}}+{\cal K}^{\bar{b} a}\nabla_a {\cal W} \nabla_{\bar{b}} {\cal \widetilde{W}}-\frac{8}{3} {\cal W} {\cal \widetilde{W}}\right) ,
\\
{\cal W} &=
\eta_1^{-2}\eta_2^{-2}\left(1+z_1z_2+z_1z_3+z_1z_4+z_2z_3+z_2z_4+z_3z_4+z_1z_2z_3z_4
\right)
\nn\\
&+
\eta_1^{-2}\eta_2^2\left(1-z_1z_2+z_1z_3-z_1z_4-z_2z_3+z_2z_4-z_3z_4+z_1z_2z_3z_4\right)
\nn\\
&+
\eta_1^4\left(1+z_1z_2-z_1z_3-z_1z_4-z_2z_3-z_2z_4+z_3z_4+z_1z_2z_3z_4\right).
\end{align}

Here $\tilde {\cal W}$ takes the same form as ${\cal W}$ except for substituting 
$z$'s for $\bar z$'s, and ${\cal K}$ is the K\"ahler potential given as
${\cal K}= - \sum_{a=1}^4 \log(1-z^a \bar{z}^{\bar{a}})$. Note also that $z$ and $\bar z$ are not complex conjugate but should be treated as two independent real scalar fields in the current context of Wick-rotated supersymmetry. With the conformal gauge choice for the metric ${\rm d}s^2 = e^{2A}(\d r^2/r^2 + \d s^2(S^4)) $ and also assuming all the scalar fields respect the symmetries of $S^4$, one may write down the BPS equations as coupled 1st-order non-linear ordinary differential equations. The explicit form of the BPS equations will be omitted here but the UV (small $r$) expansion and further numerical analysis reveal that there are six integration constants, denoted by $\mu_i,v_i, \; (i=1,2,3)$. The usual AdS/CFT dictionary tells us that $\mu_i$ are interpreted as sources for the mass terms of three chiral multiplets, while $v_i$ are the expectation values of mass terms. It turns out that, when one imposes regularity at IR (large $r$), $v_i$ are completely determined as functions of $\mu_i$. Thus the essence of the mass-deformed holographic result is encoded in the functions $v_i(\mu_j)$. 

In \cite{Bobev:2016nua}, Bobev et al. solved the BPS equations
numerically. Later in \cite{Kim:2019rwd} Kim and Kim revisited the problem semi-analytically, treating $\mu_i$ as perturbative parameter (see also \cite{Kim:2019feb,Kim:2019ewv,Kim:2020unz,Kim:2020ysx,Kim:2020dqx})  and solved the BPS equations iteratively, demanding IR regularity. The result is
\begin{align}
    v_i&=-2 \mu _i+\left(\frac{16 \pi ^4}{525}-\frac{1}{5}\right)  \mu _i^3+\left(\frac{3}{5}-\frac{8 \pi ^4}{525}\right)  \mu _i\left(\mu_1^2+\mu _2^2+\mu _3^2\right) +{\cal O}(\mu^5) . 
\end{align}
As an improvement over the analysis given in \cite{Bobev:2016nua}, where only numerical values of the expansion coefficients up to cubic order were extracted, in \cite{Kim:2019rwd} the cubic coefficients were obtained in closed form, and some higher-order coefficients were also calculated numerically up to 9th order in $\mu_i$. Given $v_i(\mu_j)$ and the supergravity solutions, it is straightforward to compute the properly-regularized on-shell action of the model \eqref{sugra5}. It turns out that the correct relation between the holography parameter and the quantum field theory parameter is $\mu_i=\pm i m_i$ \cite{Bobev:2013cja,Bobev:2016nua}. After all, the ${\cal O}(m^4)$ terms of the holographic $S^4$ free energy are given as follows \cite{Kim:2019rwd}.
\begin{align}
    \left. \frac{F}{N^2} \right|_{{\cal O}(m^4)} &=  {\cal A}(m_1^4+m_2^4+m_3^4)+{\cal B}(m_1^2+m_2^2+m_3^2)^2 . 
    \label{fn1star}
    \\
    {\cal A}&=
    +\frac{1}{40} - \frac{2\pi^4}{525} \approx -0.346082 
    \nn\\
    {\cal B}&=
    -\frac{3}{40} + \frac{\pi^4}{525} \approx +0.110541
    \nn
\end{align}

When $m_1=m_2=m$ and $m_3=0$, we have ${\cal N}=2^*$ theory. With the supergravity Lagrangian, one should set $z_2=z_4=\bar z_2=\bar z_4=0$, $w:=z_1=z_3$, $\bar w:=\bar z_1=\bar z_3$, and $\eta_1:=\eta$, $\eta_2=1$. Then the action simplifies to the one
constructed earlier in \cite{Bobev:2013cja}. The solution of the BPS equations contains two UV parameters $\mu,v$ and the authors of \cite{Bobev:2013cja} conjectured 
\begin{align}
    v = -2\mu -\mu \log (1-\mu^2) .
    \label{vmu}
\end{align}
We note that this relation has been analytically proved by identifying an 
integral of motion in \cite{Gautason:2026dcl}.
The function $v(\mu)$ is predicted naturally from the large $N$ and large 't Hooft coupling 
$\lambda:=g^2_{YM}N$ limit of the matrix model \cite{Russo:2013kea}, which is derived using the supersymmetric localization technique developed by Pestun \cite{Pestun:2007rz}. The result for our purpose is given as follows.
\begin{align}
    \frac{\d^3 F}{\d m^3} = - 2N^2\frac{m(m^2+3)}{(m^2+1)^2}.
    \label{fm3}
\end{align}
Standard prescription of AdS/CFT gives on the other hand, assuming \eqref{vmu}, $F'''(\mu)= - 2N^2 \mu (3-\mu^2)/(1-\mu^2)^2$ which agrees with \eqref{fm3} and $m=\pm i \mu$. Our 
holographic ${\cal N}=1^*$ result \eqref{fn1star} is of course consistent with the localization result, since setting $m_1=m_2=m,m_3=0$ reduces \eqref{fn1star}
to $F/N^2 |_{{\cal O}(m^4)}^{{\cal N}=2^*} = -1/4$.

Obviously the two coefficients of ${\cal O}(m^4)$ terms in \eqref{fn1star} are not 
accessible via supersymmetric localization which requires ${\cal N}=2$ supersymmetry.
One can still argue that those terms must be related to certain integrated four-point functions of scalar/spinor mass terms of ${\cal N}=1^*$ theory. In the next section we will present the computation of integrated correlators.
\section{Integrated four-point correlators}\label{sec:3}
The four-point correlators of lowest weight 1/2-BPS operators have been computed 
holographically in \cite{Arutyunov:2000py}, by considering the fluctuation mode expansion of $D=5$ supergravity and evaluating the Witten diagrams \cite{Witten:1998qj}. The so-called $\bar D$ functions, which were first defined in \cite{Arutyunov:2002fh}, prove useful in the study of four-point correlation functions in four dimensional conformal field theories. The most basic object is the box diagram, or the Bloch-Wigner function \cite{Usyukina:1993ch,Drummond:2013nda,Loebbert:2019vcj}
\begin{align}
    \bar D_{1111} &=\frac{1}{\pi^2} x^2_{13} x^2_{24}
    \int \frac{d^4 x_5}{x^2_{15}x^2_{25}x^2_{35}x^2_{45}}
    \\
     &= \frac{1}{z-\bar{z}}\,\left( \log(z\bar{z})\log(\frac{1-z}{1-\bar{z}}) + 2\text{Li}_2(z) -2\text{Li}_2(\bar{z}) \right)
     \label{bwz}
\end{align}
where $\Phi:=\bar D_{1111}$ is a function of cross ratios 
\begin{align}
    U = \frac{x^2_{12}x^2_{34}}{x^2_{13}x^2_{24}} := z\bar z
    , \quad \quad
    V = \frac{x^2_{14}x^2_{23}}{x^2_{13}x^2_{24}} := (1-z)(1-\bar z) . 
    \label{cratio}
\end{align}
Here $z,\bar z$ provide an alternative parametrization of cross ratios $U,V$ and have nothing to do with the $D=5$ supergravity scalar fields $z_a, \bar z_a$ introduced in the previous section.

According to \cite{Chester:2020dja}, the crucial integral related to ${\cal N}=2^*$ free energy in supergravity limit is given as follows.
\begin{align}
    I_4[\mathcal{T}^{SG}]
    &=\frac{1}{2\pi}  \int \d z\, \d\bar{z} (z-\bar{z})^2 \, (2+2z\bar{z}-z-\bar{z}) \overline{D}_{1111}(z,\bar{z}) \overline{D}_{2422}(z,\bar{z}) , 
    \label{ic2star}
\end{align}
where $\bar D_{2422}=\partial_{U}\partial_{V}(1+U\partial_{U}+V\partial_{V})\overline{D}_{1111}(U,V)$. This two-dimensional integral can be evaluated numerically and the result is close to $3-6\zeta(3)
\approx -4.21234$ \cite{Chester:2020dja}, 
where $\zeta(s)=\sum^\infty_{n=1} n^{-s}$ is the Riemann zeta function. $I_4[\mathcal{T}^{SG}]$ is meant to give the {\it interaction part} of the integrated correlator: the free part is computed from the one-loop determinant and for a massive hypermultiplet gives $6\zeta(3)$, exactly canceling the zeta function term of the interacting part. Although an analytic derivation of the free part is given in \cite{Chester:2020dja}, to the best of our knowledge the interacting part \eqref{ic2star} has never been computed analytically and it is our purpose to fill the gap in this paper. 

We are going to use the separation of variables, or Fourier-Mellin representation of ladder integral \cite{Loebbert:2024fsj}. 
\begin{align}
    \bar D_{1111} 
    &= \frac{1}{z-\bar{z}}\sum_{n=-\infty}^\infty \int^\infty_{-\infty} \frac{d\nu}{2\pi} \frac{n}{(\nu^2+\tfrac{n^2}{4})^2}
    (z\bar z)^{i\nu} \left( \frac{z}{\bar z} \right)^{n/2} \\
    &= \sum_{n=1}^\infty \int^\infty_{-\infty} \frac{d\nu}{2\pi} \frac{n r^{2i\nu-1}}{(\nu^2+\tfrac{n^2}{4})^2} \frac{\sin n\theta}{\sin\theta}, 
    \label{bwfm}
\end{align}
where $z=re^{i\theta}$.
One can easily verify that doing the contour integration with respect to $\nu$ gives exactly the series expansion form of \eqref{bwz}. Starting from \eqref{bwfm}, one can calculate $\bar D_{2422}(r,\theta)$, and substitute $\bar D_{1111}$ and $\bar D_{2422}$ into \eqref{ic2star}. Then one first performs the angular ($\theta$) integration, the contour integration of two $\nu$ variables for each representation of $\bar D_{1111}$ and $\bar D_{2422}$ after that. At this stage the result for $I_4[\mathcal{T}^{SG}]$ is given as $r$-integral over double sum, 
\begin{align}
    I_4[\mathcal{T}^{SG}] = \int^1_0 dr \sum_{n_1,n_2} A_{n_1,n_2}(r)
    \label{i4series}
\end{align}
where the positive integers $n_1,n_2$ are from the mode expansion of $\bar D_{2422}$ and $\bar D_{1111}$ respectively, and the radial integration range has been folded from $[0,\infty)$ to $[0,1]$ using the conformal symmetry of the correlation function. $A_{n_1,n_2}$ vanishes for $n_1=1$ or $n_2>n_1$, and we have different formulas depending on whether $n_1-n_2$ is zero, even or odd. 
The result is given as follows:
\begin{figure}[t]
    \centering
\begin{tikzpicture}
        \begin{axis}[
            axis lines = middle,
            xmin=0, xmax=14,
            ymin=0, ymax=11,
            xlabel={$n_1$},
            ylabel={$n_2$},
            xlabel style={right},
            ylabel style={above},
            xtick={1,2,3,4,5,6,7,8,9,10,11,12,13},
            ytick={1,2,3,4,5,6,7,8,9,10},
            clip=false
        ]
        
        \addplot[only marks, mark=*] coordinates {(2,1) (3,1) (4,1) (5,1) (6,1) (7,1) (8,1) (9,1) (10,1) (11,1) (12,1) (13,1)
        (13,2) (13,3) (13,4) (13,5) (13,6) (13,7) (13,8) (13,9) (13,10)
        (12,10) (11,10)
        (3,2) (4,2) (5,2) (6,2) (7,2) (8,2) (9,2) (10,2) (11,2) (12,2)
         (4,3) (5,3) (6,3) (7,3) (8,3) (9,3) (10,3) (11,3) (12,3)
         (5,4) (6,4) (7,4) (8,4) (9,4) (10,4) (11,4) (12,4)
         (6,5) (7,5) (8,5) (9,5) (10,5) (11,5) (12,5)
         (7,6) (8,6) (9,6) (10,6) (11,6) (12,6)
        (8,7) (9,7) (10,7) (11,7) (12,7)
         (9,8) (10,8) (11,8) (12,8)
         (10,9) (11,9) (12,9) (2,2)(3,3)(4,4)(5,5)(6,6)(7,7)(8,8)(9,9)(10,10)};

         \addplot[only marks, mark=*, mark options={fill=purple, draw=purple}] 
            coordinates {(2,2) (3,3)(4,4) (5,5)(6,6)(7,7)(8,8)(9,9)(10,10)};
            \addplot[only marks, mark=*, mark options={fill=blue, draw=blue}] 
            coordinates {(2,1)(3,2)(4,3)(5,4)(6,5)(7,6)(8,7)(9,8)(10,9)(11,10)(4,1)(5,2)(6,3)(7,4)(8,5)(9,6)(10,7)(11,8)(12,9)(13,10)(6,1)(7,2)(8,3)(9,4)(10,5)(11,6)(12,7)(13,8)(8,1)(9,2)(10,3)(11,4)(12,5)(13,6)(10,1)(11,2)(12,3)(13,4)(12,1)(13,2)};
            \addplot[only marks, mark=*, mark options={fill=red, draw=red}] 
            coordinates {(3,1)(5,1)(7,1)(9,1)(11,1)(13,1)(4,2)(6,2)(8,2)(10,2)(12,2)(5,3)(7,3)(9,3)(11,3)(13,3)(6,4)(8,4)(10,4)(12,4)(7,5)(9,5)(11,5)(13,5)(8,6)(10,6)(12,6)(9,7)(11,7)(13,7)(10,8)(12,8)(11,9)(13,9)(12,10)};

        \draw[thick,->,red] 
        (axis cs:2,2) -- (axis cs:11,11);

        \node[anchor=west] at (axis cs:11.5,11.5) {$\pi^2 - 6\zeta(3)$};

        \end{axis}
    \end{tikzpicture}
     \caption{Formulas \eqref{Adiag}, \eqref{Aeven}, and \eqref{Aodd} cover the $(n_1,n_2)$ lattice points corresponding to the purple, red, and blue dots, respectively.}
    \label{fig:n2sum1}
\end{figure}
\small
\begin{align}
    A_{n_1=n_2}(r) &=\frac{8 r^{2 n_2-1} }{n_2^2}\left(1-n_2 \log (r)\right)\left(n_2-1\right)\,,\label{Adiag}
\end{align}
\begin{align}
    A_{n_1-n_2 \in 2{\mathbb{Z}_+}}(r)&=\frac{r^{n_1+n_2-3}}{n_1^2 n_2}(n_2 \log(r)-1)\Big[\left(-n_1^4+2 n_2^2 n_1^2-4 n_1^2-n_2^4+4 n_2^2\right) r^4\nn\\
    &-4 \left(n_1^4-2 n_2^2 n_1^2+2 n_1^2+n_2^4+2 n_2^2\right) r^2 +\left(-n_1^4-n_2^4-4 n_1^2+2 n_1^2 n_2^2+4 n_2^2\right)\nn\\
    &+\left(n_1^5+4 n_1^4-2 n_2^2 n_1^3+4 n_1^3-4 n_2^2 n_1^2+n_2^4 n_1-4 n_2^2 n_1\right) r^4 \log(r)\nn\\
    &+4 \left(n_1^5-2 n_2^2 n_1^3-2 n_1^3+n_2^4 n_1+2 n_2^2 n_1\right)
   r^2 \log(r)\nn\\
   &+\left(n_1^5-4 n_1^4-2 n_2^2 n_1^3+4 n_1^3+4 n_2^2 n_1^2+n_2^4 n_1-4 n_2^2 n_1 \right)\log(r) \Big]\label{Aeven}
\end{align}
\begin{align}
   A_{n_1-n_2 \in 2{\mathbb{Z}_+}-1}(r) &= \frac{r^{n_1+n_2-2}}{n_1^2 n_2}(n_2 \log(r)-1)\Big[ \left(3 n_1^4-6 n_2^2 n_1^2+10 n_1^2+3 n_2^4+2 n_2^2-5\right) r^2\nn\\
   &+\left(3 n_1^4-6 n_2^2 n_1^2+10 n_1^2+3 n_2^4+2 n_2^2-5\right)\nn\\
   &+\left(-3 n_1^5-4 n_1^4+6 n_2^2 n_1^3+6 n_1^3+4 n_2^2 n_1^2+12 n_1^2-3 n_2^4 n_1-2 n_2^2 n_1+5 n_1\right) r^2 \log(r)\nn\\
   &+\left(-3 n_1^5+4 n_1^4+6 n_2^2 n_1^3+6 n_1^3-4 n_2^2 n_1^2-12
   n_1^2-3 n_2^4 n_1-2 n_2^2 n_1+5 n_1\right)\log(r) \Big]\,.\label{Aodd}
\end{align}\normalsize

One can easily check $A_{n_1,n_2}(r)$ is finite-valued in $0\le r\le 1$,
and $\int^{1}_{0} dr A_{n_1,n_2}$ always turns out to be a rational number. 
$\int^1_0 dr A_{n_1,n_2}$ for a few small values of $n_1,n_2$ is given in the table below.
\begin{center}
\begin{tabular}{|c|c|c|c|c|c|c|c|c|}
      \hline
      $n_2=8$ & 0 & 0 & 0 & 0 & 0 & 0 & 0& $\frac{21}{256}$  \\
      \hline
      $n_2=7$ & 0 & 0 & 0 & 0 & 0 & 0 & $\frac{36}{343}$& -$\frac{4513}{6272}$  \\
      \hline
      $n_2=6$ & 0 & 0 & 0 & 0 & 0 & $\frac{5}{36}$ & -$\frac{1951}{2058}$& $\frac{14853}{6272}$  \\
      \hline
      $n_2=5$ & 0 & 0 & 0 & 0 & $\frac{24}{125}$& -$\frac{98}{75}$& $\frac{245177}{77175}$& -$\frac{100181}{17640}$  \\
      \hline
      $n_2=4$ & 0 & 0 & 0 & $\frac{9}{32}$& -$\frac{1919}{1000}$& $\frac{2699}{600}$& -$\frac{86258}{11025}$& $\frac{497573}{44100}$  \\
      \hline
      $n_2=3$ & 0 & 0 & $\frac{4}{9}$& -$\frac{149}{48}$& $\frac{10361}{1500}$& -$\frac{1157}{100}$& $\frac{179371}{11025}$& -$\frac{4738}{225}$  \\
      \hline
      $n_2=2$ & 0 & $\frac{3}{4}$ & -$\frac{107}{18}$ & $\frac{1757}{144}$& -$\frac{34753}{1800}$& $\frac{70957}{2700}$& -$\frac{325249}{9800}$& $\frac{1997}{50}$  \\
      \hline
      $n_2=1$ & 0 & -17 & $\frac{271}{9}$& -$\frac{1529}{36}$& $\frac{12479}{225}$& -$\frac{3665}{54}$& $\frac{394087}{4900}$& -$\frac{2318}{25}$  \\
      \hline
       & $n_1=1$ & $n_1=2$ & $n_1=3$&$n_1=4$&$n_1=5$&$n_1=6$&$n_1=7$&$n_1=8$\\
      \hline
    \end{tabular}
\end{center}

\begin{figure}[h]
    \centering
    \includegraphics[width=0.5\textwidth]{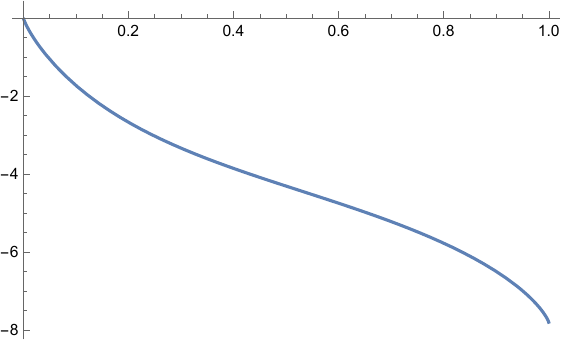}
    \caption{Integrand of $I_4[\mathcal{T}^{SG}]$}
    \label{I4TSGgraph}
\end{figure}

We  note that this is a divergent series: $\lim_{n_1\rightarrow\infty}|\int^1_0 dr A_{n_1,n_2}|=\infty$. One may try to re-sum
the series using Abel's summation method\footnote{A simple and motivational example for us would be   $ 1-2+3-4+\cdots =\lim_{t\rightarrow 1^-} (t-2t^2+3t^3-4t^4+\cdots)=\lim_{t\rightarrow 1^-} \left[ \frac{t}{(1+t)^2} \right]=\frac{1}{4}$. This can be considered as an alternative way of evaluating $I=\int^1_0 \frac{1-x}{(1+x)^3} dx=\frac{1}{4}$, which is a completely harmless integration. When we series expand the integrand however, $I=\int^1_0 1- 4x +9 x^2 -16 x^3 +\cdots = 1-2+3-4+\cdots$. For this example we can of course trivially do the integral, but this prescription can be applied to non-trivial integrals as well. The problem here derives from the fact that the radius of convergence for the series is one and the range of integration touches it. It is also the case with \eqref{ic2star}, since the summation in \eqref{bwfm} breaks down at $|z|=1$.}. However, since we are given a double series, there is an issue of choosing the right ordering of the summation and the result in general does depend on the ordering of our choice. We thus do the summation of the integrands first and
then perform the integral after that. This procedure poses no ambiguity,
since the double sum over $A_{n_1,n_2}$ is convergent at every point in $0<r<1$, and the result of the summation behaves well in $0\le r \le 1$. The summed integrand is plotted in Fig.\ref{I4TSGgraph}.

For practical computation, we split the sum into three parts.
Firstly, along the line $n_1=n_2$, the single sum and the $r$-integration can be performed easily, yielding $\int^1_0 dr \sum^\infty_{n_2=2} A_{n_1=n_2}=\pi^2-6\zeta(3)$. Then we turn to $n_1\neq n_2$, and consider even and odd $n_1-n_2$ cases separately. It is obviously useful to introduce a new summation index from $n_1-n_2$. In general the summands contain the following expressions.
\begin{align}
    &\sum^{\infty}_{d_1=1} \sum^{\infty}_{n_2=1}(2 d_1 + n_2)^{\alpha}\, n_2^{\beta}\, r^{2 d_1 + 2 n_2 - \gamma} \quad,\quad 2 d_1=n_1-n_2\\
    &\sum^{\infty}_{d_2=1} \sum^{\infty}_{n_2=1}(2 d_2 + n_2-1)^{\alpha}\, n_2^{\beta}\, r^{2 d_2 + 2 n_2 -1 - \gamma} \quad,\quad 2 d_2 -1=n_1-n_2
\end{align}
where each expression comes from $\sum A_{n_1-n_2 \in 2{\mathbb{Z}_+}}(r)$ and $\sum A_{n_1-n_2 \in 2{\mathbb{Z}_+}-1}(r)$ respectively. After carrying out the double sums and simplifying with standard polylogarithm identities, the final integrand takes the schematic form
\begin{align}
    \frac{\text{product of\,} \left\{ \, \log(\cdots),\text{Li}_2(\cdots),\text{Li}_3(\cdots) \, \right\} }{(\cdots)^k}
    \label{prodOflogs}
\end{align}
where each ($\cdots$) is one of $\{r\,,\,1-r\,,\,1+r\}$ and $k\in \mathbb{Z}_{+}$.
The total poly-logarithmic weight in the numerator does not exceed three (we assign weight 1 on $\log$ and $n$ on $\text{Li}_n (r)$).
When we consider \eqref{ic2star}, the integrals of the above type are readily done by indefinite integration, and evaluating between the endpoints gives $I_4[{\cal T}^{SG}]=3-6\zeta(3)$ analytically.

Let us now turn to the computation of integrated correlators which are related to the ${\cal O}(m^4)$ terms of ${\cal N}=1^*$ theories. We closely follow and make use of the results given in \cite{Chester:2020dja}, so the readers are referred to this reference for more details. When we give mass to three chiral multiplets of ${\cal N}=4$ super Yang-Mills theory on $S^4$, we need to add three kinds of new terms to the action.
\begin{align}
    J_{i}\equiv \frac 12  \mathrm{tr} \left( Z_i^2 + \bar Z_i^2 \right) \,, \qquad
   K_{i} \equiv - \frac 12 \mathrm{tr} (\chi_i \sigma_2 \chi_i + \tilde \chi_i \sigma_2 \tilde \chi_i) \,, \qquad {L_i \equiv \mathrm{tr}  |{Z_i}|^2  } \,.
\end{align}
where $Z_i,\bar Z_i, \;i=1,2,3$ are three complex scalars, and $\chi_i,\tilde \chi_i$ are their superpartner spinors.
The mass-dependent part of the action is, 
\begin{align}
    S_m = \sum_{i=1}^3\int {\d}^4x \sqrt{g}\left[ m_i(iJ_i+K_i)+m^2_iL_i\right] .
\end{align}
Since we are interested in the {\it interaction} part of the four-point correlators which contribute to the {\it quartic} order terms of the free energy, we may ignore $L_i$ terms from now on. Thanks to the $SU(4)_R$ symmetry, the free energy as a function of the mass parameters must be symmetric under permutations of $m_i$. For quartic terms, we thus need to evaluate only two numbers, e.g.
\begin{align}
      \mathcal{F}_{1111} \equiv \frac{\partial^4 F(m_1,m_2,m_3)}{\partial m_{1}^4} \bigg\vert_{m_{i}=0} , \\
      \mathcal{F}_{1122} \equiv \frac{\partial^4 F(m_1,m_2,m_3)}{\partial m_{1}^2\partial m_{2}^2} \bigg\vert_{m_{i}=0} . 
\end{align}

The scalar quadratic operators $Z_i^2$ can be re-expressed in an $SO(6)_R$ symmetric fashion, $S(Y)=\mathrm{tr}(\phi_I\phi_J)Y^IY^J$. Here $\phi_I, \; I=1,\cdots,6$ are six real scalars which make $Z_1=\phi_1+i\phi_2$ etc, and $Y^I$ is a six-dimensional null vector which describes the $SO(6)$ polarization of the specific 1/2-BPS operator $S(Y)$. The 4-point correlation function of 1/2-BPS operators exhibits a crucial property of factorization. In particular, 
\begin{align}\langle S(x_1,Y_1)S(x_2,Y_2)S(x_3,Y_3)S(x_4,Y_4)\rangle
= \frac{1}{x_{12}^4 x^4_{34}} \vec{\cal S}\cdot \vec{\cal B}
\label{sugrat}
\end{align}
where ${\cal S}$ encodes the spatial dependence and splits into free part and interaction part, i.e. $\vec{\cal S}=\vec{\cal S}_{\rm free}+ \vec{\bf S}{\cal T}$. Here $\vec{\cal S}_{\rm free}, \vec{\bf S}$ are six different functions of the cross ratios $U,V$ and they are universal in the sense that they are independent of $g_{YM}$ and $N$. ${\cal T}(U,V)$ on the other hand has a non-trivial dependence on $g_{YM},N$, so it contains the non-trivial physics of interacting ${\cal N}=4$ super Yang-Mills, but it is universal in the sense that it is independent of $SO(6)_R$ polarization vectors $Y_I$. The AdS/CFT calculation using $D=5$ gauged supergravity gives the strongly coupled 't Hooft limit answer \cite{Goncalves:2014ffa,Arutyunov:2000py}.
\begin{align}
    {\cal T}^{\rm SG} = - \frac{1}{8} U^2 \bar D_{2422} . 
    \label{sugt}
\end{align}

We need to consider spinor mass terms as well. The spinors of ${\cal N}=4$ super Yang-Mills are in ${\bf 4}$ of $SU(4)_R$, so it is useful to introduce $SU(4)_R$ polarization vectors $X_A$ and $\bar X^A$. Their 4-point functions take the following form,
\begin{align}
    \langle P(x_1,\bar X_1)\bar P(x_2,X_2) P(x_3,\bar X_3)\bar P(x_4,X_4)\rangle
= \frac{1}{x_{12}^6 x^6_{34}} \vec{\cal P}\cdot \vec{\cal B}_P . 
\end{align}
And there are also contributions from the mixed correlator of scalars and spinors,
\begin{align}
    \langle S(x_1,Y_1) S(x_2,Y_2) \bar P(x_3,\bar X_3) P(x_4,X_4)\rangle
= \frac{1}{x_{12}^4 x^6_{34}} \vec{\cal R}\cdot \vec{\cal B}_{SP} .
\end{align}
In a similar fashion to scalar 4-point functions, $\vec {\cal P}$ and $\vec {\cal R}$, now 3 dimensional vectors, split into the free and the interaction part.
\begin{align}
    \vec{\cal P} (U,V) &= \vec{\cal P}_{\rm free}(U,V) + \vec{\bf P}(U,V,\partial_U,\partial_V) {\cal T}(U,V) \\
    \vec{\cal R} (U,V) &= \vec{\cal R}_{\rm free}(U,V) + \vec{\bf R}(U,V,\partial_U,\partial_V) {\cal T}(U,V) 
    . 
\end{align}
The three kinds of 4-point functions are closely related via Ward identities and in terms of a common amplitude ${\cal T}$. $\vec{\bf P},\vec{\bf R}$ are independent of coupling constants, and their concrete forms can be found in \cite{Chester:2020dja}.

\begin{figure}[h]
    \centering
    \includegraphics[width=0.5\textwidth]{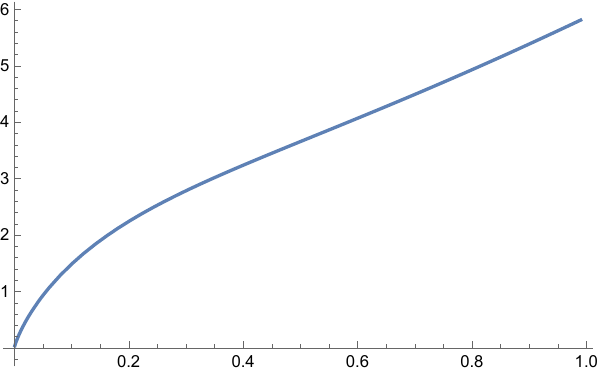}
    \caption{Integrand of $\frac{1}{c}{\cal F}_{1122}^{\rm SG}$}
    \label{F1122graph}
\end{figure}

We can now write down the expressions for ${\cal F}_{1111},{\cal F}_{1122}$ in a similar way as the ${\cal N}=2^*$ result \eqref{ic2star}. 
\begin{align}
    {\cal F}_{1111}
    &= -\frac{c^2}{64 \pi^8}  I_{2, 2}^4 \left[ \big(4 \left( {\bf S}_1 + {\bf S}_2 + {\bf S}_3 \right) + 2 ({\bf S}_4 + {\bf S}_5 + {\bf S}_6 ) \big) {\cal T} \right]
    \nn\\
    &+  \frac{ c^2}{ 8 \pi^8} I_{2, 3}^4 \left[24 ({\bf R}_1 + {\bf R}_3){\cal T} \right] 
    - \frac{c^2}{\pi^8} I_{3, 3}^4 \left[ 6 ({\bf P}_1 + {\bf P}_2 + {\bf P}_3) {\cal T} \right] \\
    {\cal F}_{1122}
    &=-\frac{c^2}{64 \pi^8}  I_{2, 2}^4 \left[ \frac{4}{3} \left( {\bf S}_1 + {\bf S}_2 + {\bf S}_3 \right)  {\cal T} \right]+  \frac{ c^2}{ 8 \pi^8} I_{2, 3}^4 \left[ (8{\bf R}_1 + 40{\bf R}_3){\cal T} \right]\nn \\
    &- \frac{c^2}{\pi^8} I_{3, 3}^4 \left[ 2 ({\bf P}_1 + {\bf P}_2 ) {\cal T} \right]
\end{align}
Here $c=(N^2-1)/4$, and with $I^4_{2,2}[\bullet],I^4_{2,3}[\bullet],I^{4}_{3,3}[\bullet]$ one is instructed to do the following spatial integration.
\begin{align}
    I^{4}_{\Delta_A,\Delta_B} [{\cal G}] &= \frac{2^{17-2\Delta_A-2\Delta_B}\pi^7\Gamma(6-\Delta_A-\Delta_B)}{3\Gamma(4-\Delta_A)^2\Gamma(4-\Delta_B)^2}\nn\\
    &\times \int \d r \d \theta \, r^3\sin^2\theta \, 
    \bar D_{4-\Delta_A,4-\Delta_A,4-\Delta_B,4-\Delta_B}(U,V) \frac{\cal G}{U^{\Delta_A}} 
\end{align}
with $r,\theta$ parametrized as in \cite{Chester:2020dja}.
One can proceed with the computation, and just as in \cite{Chester:2020dja}, it is convenient to make use of the Mellin transform of $\bar D$ functions. After rather lengthy calculation using the explicit form of $\vec{\bf P},\vec{\bf R}$, we arrive at the following results.
\begin{align}
    {\cal F}_{1111}^{\rm SG} &= - \frac{3c}{\pi}\,\int \d z\,\d\bar{z}\,(z-\bar{z})^2 \overline{D}_{2422} (1+z\bar{z})\left(\overline{D}_{1111}-2 \overline{D}_{2112}\right), \\
    {\cal F}_{1122}^{\rm SG} &= -\frac{2c}{\pi}\,\int \d z\,\d \bar{z}\, (z-\bar{z})^2 \,\overline{D}_{2422} (1+z\bar{z})\overline{D}_{2112} .
\end{align}
Since $\frac{1}{c}{\cal F}_{1111}^{\rm SG}+\frac{3}{c}{\cal F}_{1122}^{\rm SG}=-4 \, I_4[\mathcal{T}^{SG}]$ and we have already done the ${\cal N}=2^*$ computation, we need to compute only one of them and obviously for simplicity we choose to work on ${\cal F}_{1122}^{\rm SG}$. Again using the Fourier-Mellin representation \eqref{bwfm}, this integral is given as $r$-integral over double series $\int^1_0 dr \sum_{n_1,n_2} C_{n_1,n_2}(r)$. The formula for $C_{n_1,n_2}(r)$ is lengthy and relegated to Appendix~\ref{app:F1122}.
\begin{figure}[t]
    \centering
\begin{tikzpicture}
        \begin{axis}[
            axis lines = middle,
            xmin=0, xmax=14,
            ymin=0, ymax=11,
            xlabel={$n_1$},
            ylabel={$n_2$},
            xlabel style={right},
            ylabel style={above},
            xtick={1,2,3,4,5,6,7,8,9,10,11,12,13},
            ytick={1,2,3,4,5,6,7,8,9,10},
            clip=false
        ]
        
        \addplot[only marks, mark=*] coordinates { (13,2) (13,3) (13,4) (13,5) (13,6) (13,7) (13,8) (13,9) (13,10)
        (12,10) (11,10)
        (3,2) (4,2) (5,2) (6,2) (7,2) (8,2) (9,2) (10,2) (11,2) (12,2)
         (4,3) (5,3) (6,3) (7,3) (8,3) (9,3) (10,3) (11,3) (12,3)
         (5,4) (6,4) (7,4) (8,4) (9,4) (10,4) (11,4) (12,4)
         (6,5) (7,5) (8,5) (9,5) (10,5) (11,5) (12,5)
         (7,6) (8,6) (9,6) (10,6) (11,6) (12,6)
        (8,7) (9,7) (10,7) (11,7) (12,7)
         (9,8) (10,8) (11,8) (12,8)
         (10,9) (11,9) (12,9) (2,2)(3,3)(4,4)(5,5)(6,6)(7,7)(8,8)(9,9)(10,10) (2,3) (2,4)(2,5)(2,6)(2,7)(2,8)(2,9)(2,10)(3,4)(3,5)(3,6)(3,7)(3,8)(3,9)(3,10)(4,5)(4,6)(4,7)(4,8)(4,9)(4,10)(5,6)(5,7)(5,8)(5,9)(5,10)(6,7)(6,8)(6,9)(6,10)(7,8)(7,9)(7,10)(8,9)(8,10)(9,10) };

            \addplot[only marks, mark=*, mark options={fill=blue, draw=blue}] 
            coordinates {(3,2)(4,3)(5,4)(6,5)(7,6)(8,7)(9,8)(10,9)(11,10)(5,2)(6,3)(7,4)(8,5)(9,6)(10,7)(11,8)(12,9)(13,10)(7,2)(8,3)(9,4)(10,5)(11,6)(12,7)(13,8)(9,2)(10,3)(11,4)(12,5)(13,6)(11,2)(12,3)(13,4)(13,2) (2,3)(2,5)(2,7)(2,9)(3,4)(3,6)(3,8)(3,10)(4,5)(4,7)(4,9)(5,6)(5,8)(5,10)(6,7)(6,9)(7,8)(7,10)(8,9)(9,10)};
            \addplot[only marks, mark=*, mark options={fill=red, draw=red}] 
            coordinates {(4,2)(6,2)(8,2)(10,2)(12,2)(5,3)(7,3)(9,3)(11,3)(13,3)(6,4)(8,4)(10,4)(12,4)(7,5)(9,5)(11,5)(13,5)(8,6)(10,6)(12,6)(9,7)(11,7)(13,7)(10,8)(12,8)(11,9)(13,9)(12,10)(2,2) (3,3)(4,4) (5,5)(6,6)(7,7)(8,8)(9,9)(10,10) (2,4)(2,6)(2,8)(2,10)(3,5)(3,7)(3,9)(4,6)(4,8)(4,10)(5,7)(5,9)(6,8)(6,10)(7,9)(8,10)};
        
        \end{axis}
    \end{tikzpicture}
    \caption{Formulas \eqref{Ceven-le} and \eqref{Ceven-ge} cover the $(n_1,n_2)$ lattice points shown as red dots, while \eqref{Codd-le} and \eqref{Codd-ge} cover those shown as blue dots.}
    \label{fig::n1sum1}
\end{figure}
We find $C_{n_1,n_2}$ is non-zero for $n_1,n_2\ge 2$, as shown 
in Fig.\ref{fig::n1sum1}. Below we provide a table made of some representative values of each integral.
\begin{center}
\begin{small}
\begin{tabular}{|c|c|c|c|c|c|c|c|c|}
      \hline
      $n_2=8$ & 0 & $\frac{337}{900}$ & -$\frac{8584361}{4630500}$ & $\frac{3008281}{514500}$ & -$\frac{128942203}{8820000}$ & $\frac{6218251}{197568}$ & -$\frac{3854111}{63504}$ & $\frac{382558273}{3556224}$  \\
      \hline
      $n_2=7$ & 0 & -$\frac{647}{1225}$ & $\frac{440576}{165375}$ & -$\frac{4977793}{600250}$ & $\frac{278748712}{13505625}$ & -$\frac{29213983}{661500}$ & $\frac{51243637}{605052}$& -$\frac{1504547}{10584}$  \\
      \hline
      $n_2=6$ & 0 & $\frac{1117}{1350}$ & -$\frac{161311}{40500}$ & $\frac{928}{75}$ & -$\frac{47096687}{1543500}$ & $\frac{225206314}{3472875}$ & -$\frac{101493583}{882000}$& $\frac{34851793}{197568}$  \\
      \hline
      $n_2=5$ & 0 & -$\frac{293}{225}$ & $\frac{108811}{16875}$ & -$\frac{8837}{450}$ & $\frac{90053}{1875}$& -$\frac{21037706}{231525}$& $\frac{1933402888}{13505625}$& -$\frac{176125367}{882000}$  \\
      \hline
      $n_2=4$ & 0 & $\frac{179}{72}$ & -$\frac{152839}{13500}$ & $\frac{34219}{1000}$& -$\frac{312299}{4500}$& $\frac{30199}{270}$& -$\frac{565559713}{3601500}$& $\frac{20861311}{102900}$  \\
      \hline
      $n_2=3$ & 0 & -$\frac{44}{9}$ & $\frac{1895}{81}$& -$\frac{113317}{2250}$& $\frac{152891}{1875}$& -$\frac{1154954}{10125}$& $\frac{3453368}{23625}$& -$\frac{82543387}{463050}$  \\
      \hline
      $n_2=2$ & 0 & $\frac{142}{9}$ & -$\frac{3563}{108}$ & $\frac{3563}{72}$& -$\frac{1513801}{22500}$& $\frac{568079}{6750}$& -$\frac{11170183}{110250}$& $\frac{42491}{360}$  \\
      \hline
      $n_2=1$ & 0 & 0 & 0& 0& 0& 0& 0& 0  \\
      \hline
       & $n_1=1$ & $n_1=2$ & $n_1=3$&$n_1=4$&$n_1=5$&$n_1=6$&$n_1=7$&$n_1=8$\\
      \hline
    \end{tabular}
    \end{small}
\end{center}
As in the $\mathcal{N}=2^*$ case, performing the double summation first yields the $r$-integrand of the type in \eqref{prodOflogs}. Its plot
is given in Fig.\ref{F1122graph}. 
Unlike $I_4[{\cal T}^{SG}]$, we find indefinite integration is not available for several terms in the integrand. Fortunately, definite integrals can be obtained for such terms as well, in the following way. We use the series representations of the polylogarithms $\text{Li}_s(z)=\sum_{k=1}^\infty\frac{z^k}{k^s}$, and then do the integral term-by-term, thus rewriting the integral as a sum. Applying Abel's method to those series gives the answer.
Finally, the integral is evaluated to be
\begin{align}
   -\frac{2}{\pi}\,\int dz\,d \bar{z}\, (z-\bar{z})^2 \,\overline{D}_{2422} (1+z\bar{z})\overline{D}_{2112}  = -\frac{12}{5} + \frac{32\pi^4}{525} . 
\end{align}
To compare with the supergravity result \eqref{fn1star}, this number is predicted to be $32{\cal B}$ and we have an exact match. Note that this number is related to mixed derivatives $\partial^4{\cal F}/\partial_{m_1}^2\partial_{m_2}^2$, which means there is no contribution from the free part.

\section{Discussions}\label{sec:4}
In this paper we have provided a vindication of the supergravity computation for mass-deformed ${\cal N}=4$ super Yang-Mills theory on $S^4$. The coefficients at quartic order, ${\cal A,B}$ in \eqref{fn1star}, have remained mysterious but we have verified them as integrated 4-point correlation functions.  We should mention that this is {\it not} a non-trivial check of AdS/CFT in the usual sense, i.e. comparison of a genuinely quantum field theory calculation against a classical supergravity calculation. For ${\cal N}=2^*$, \eqref{fm3} is derivable both using supersymmetric localization and classical supergravity, so it is a feat of AdS/CFT. On the other hand, because localization formula is not available for ${\cal N}=1^*$, our computation in this paper is based on a supergravity result given in \cite{Arutyunov:2000py}, compared to another supergravity result obtained in \cite{Kim:2019rwd}. The difference between these two works is that the authors of \cite{Arutyunov:2000py} used Witten diagrams to calculate 4-point functions holographically, while the approach taken in \cite{Bobev:2013cja,Bobev:2016nua} etc. is to construct the whole supergravity solutions when one turns on the dual of mass operators. In the former approach one keeps the information of all the subtle functional dependence on cross ratios, but higher-point functions have to be computed one-by-one. In the latter approach the spatial, or cross ratio dependence is swept away, but potentially the on-shell action gives information on all higher-point correlation functions - integrated with appropriate measures - in one go.

Admittedly, the calculation presented in this work looks like quite an inefficient way to arrive at a single number. One might try to do the integral using Mellin transform instead. In fact, while the supergravity level amplitude ${\cal T}^{\rm SG}=-U^2\bar D_{2422}/8$ \eqref{sugt} is not everyone's first guess, in Mellin representation it is probably more  appealing aesthetically, being simply ${\cal M}^{\rm SG}=1/(stu)$. Indeed, the authors of \cite{Chester:2020dja} computed the 1st integrated correlators using the second Barnes' lemma, without much pain. It is straightforward to write down the Mellin form of the integrals we have evaluated in this work, but as it turns out they are not directly amenable to the use of Barnes' lemma. One might pursue along the line of \cite{Jantzen:2012cb} to derive more intricate Barnes-like lemmas. Certainly, any other method which streamlines the evaluation of the 2nd integrated correlators will be desirable. For the time being, the method
proposed in this article can be applied to other integrated correlators 
in $D=4$ conformal field theories. We are currently working on the evaluation of sub-leading terms of integrated correlator given in Eq.(2.17) of \cite{Chester:2020dja} and the result will be reported separately.

\acknowledgments
We thank Jung-Wook Kim, Congkao Wen, and Shun-Qing Zhang for discussions.
This work is supported by the National Research Foundation of Korea under the grant NRF-2022R1A2B5B02002247. NK would like to thank the organizers and speakers 
of the iTHEMS-YITP Workshop: Bootstrap, Localization and Holography (20th-24th May 2024), where he learned a lot about integrated correlators.

\appendix
\section[The calculation of a mixed quartic-order coefficient]{The calculation of ${\cal F}_{1122}$}\label{app:F1122}
Here we record the mode expansion integration result for 
\begin{align}
    -\frac{2}{\pi}\,\int dz\,d \bar{z}\, (z-\bar{z})^2 \,\overline{D}_{2422} (1+z\bar{z})\overline{D}_{2112} = \int^1_0 dr \sum_{n_1,n_2} C_{n_1,n_2}(r) .
\end{align}
The mode numbers $n_1,n_2$ are for the Fourier-Mellin expansion of $\bar D_{2422},\bar D_{2112}$ respectively. It turns out we have 4 different formulas: depending on whether $n_1\ge n_2$ or $n_2\ge n_1$, and whether $n_1-n_2$ is even or odd. 

\begin{scriptsize}
\begin{align}
    C^{n_1-n_2 \in 2{\mathbb{Z}}}_{n_1 \leq n_2}(r)=&\frac{r^{n_1+n_2-3}}{105 n_1 n_2^2}(1+r^2)(1-n_1^2)(n_2 \log(r)-1)\Big\{ \left(32 n_1^4+480 n_1^2-192\right) r^2
    \nn\\&+(32 n_1^4+480 n_1^2-192)
    +\left(-32 n_1^5-112 n_1^4+80 n_1^3+448 n_1^2+192 n_1\right) r^2 \log(r)
    \nn\\&+(-32 n_1^5+112 n_1^4+80 n_1^3-448 n_1^2+192 n_1)\log(r) \Big\}\,.\label{Ceven-le}
   \end{align}
   \begin{align}
   C^{n_1-n_2 \in 2{\mathbb{Z} +1}}_{n_1 \leq n_2}(r)=&\frac{r^{n_1+n_2-4}}{105 n_1 n_2^2}(1+r^2)(1-n_1^2)(n_2 \log(r)-1)\Big\{ \left(-8 n_1^4-64 n_1^2+384\right) r^4
   +\left(-48 n_1^4-832 n_1^2-384\right) r^2
   \nn\\&+(-8 n_1^4-64 n_1^2+384)
   +\left(8 n_1^5+56 n_1^4+64 n_1^3-224 n_1^2-384 n_1\right) r^4\log(r)
   \nn\\&+\left(48 n_1^5-288 n_1^3+384 n_1\right) r^2\log(r)
   +(8 n_1^5-56 n_1^4+64 n_1^3+224 n_1^2-384 n_1)\log(r) \Big\}\,.\label{Codd-le}
\end{align}

\begin{align}
    C^{n_1-n_2 \in 2{\mathbb{Z}}}_{n_1 \geq n_2}(r)=&\frac{r^{n_1+n_2-3}}{105 n_1^2 n_2}(1+r^2)(1-n_2^2)(n_2 \log(r)-1)\Big\{ 4 \left(35 n_1^4-42 n_2^2 n_1^2+168 n_1^2+15 n_2^4-48 n_2^2-48\right) r^2
    \nn\\&+4 \left(35 n_1^4-42 n_2^2 n_1^2+168 n_1^2+15 n_2^4-48 n_2^2-48\right)
    \nn\\&+(-140 n_1^5-280 n_1^4+168 n_2^2 n_1^3-112 n_1^3+168 n_2^2 n_1^2+448 n_1^2-60 n_2^4 n_1+192 n_2^2 n_1+192 n_1)r^2 \log(r)
    \nn\\&+(-140 n_1^5+280 n_1^4+168 n_2^2 n_1^3-112 n_1^3-168 n_2^2 n_1^2-448 n_1^2-60 n_2^4 n_1+192 n_2^2 n_1+192 n_1)\log(r) \Big\}\,.\label{Ceven-ge}
   \end{align}
   \begin{align}
    C^{n_1-n_2 \in 2{\mathbb{Z} +1}}_{n_1 \geq n_2}(r)=&\frac{r^{n_1+n_2-4}}{105 n_1^2 n_2}(1+r^2)(1-n_2^2)(n_2 \log(r)-1)\Big\{ \left(-35 n_1^4+42 n_2^2 n_1^2-238 n_1^2-15 n_2^4+174 n_2^2-351\right) r^4
    \nn\\&+\left(-210 n_1^4+252 n_2^2 n_1^2-868 n_1^2-90 n_2^4+36 n_2^2+246\right) r^2
    \nn\\&+(-35 n_1^4-15 n_2^4-238n_1^2+42 n_1^2 n_2^2+174 n_2^2-351)
    \nn\\&+\left(35 n_1^5+140 n_1^4-42 n_2^2 n_1^3+238 n_1^3-84 n_2^2 n_1^2+196 n_1^2+15 n_2^4 n_1-174 n_2^2 n_1+351 n_1\right) r^4\log(r)
    \nn\\&+\left(210 n_1^5-252 n_2^2 n_1^3-252 n_1^3+90 n_2^4
   n_1-36 n_2^2 n_1-246 n_1\right) r^2\log(r)
   \nn\\&+(35 n_1^5-140 n_1^4-42 n_2^2 n_1^3+238 n_1^3+84 n_2^2 n_1^2-196 n_1^2+15 n_2^4 n_1
   \nn\\&-174 n_2^2 n_1+351 n_1)\log(r) \Big\}\,.\label{Codd-ge}
\end{align}
\end{scriptsize}

\bibliographystyle{JHEP}
\bibliography{biblio.bib}
\end{document}